\documentclass[aps,prc,twocolumn,superscriptaddress,showpacs,showkeys]{revtex4}
\usepackage{hypernat}
\usepackage{hyperref}

\usepackage{graphicx}% Include figure files
\usepackage{dcolumn}% Align table columns on decimal point
\usepackage{bm}% bold math

\usepackage{float}

\usepackage[usenames]{color}
\begin{document}

%\preprint{}

\title{Non-equilibrium emission of complex fragments from p+Au collisions at 2.5 GeV
proton beam energy}


\author{A.Bubak}
%\thanks{}
\affiliation{Institut f{\"u}r Kernphysik, Forschungszentrum
J{\"u}lich, D-52425 J{\"u}lich, Germany} \affiliation{Institute of
Physics, Silesian University, Uniwersytecka 4, 40007 Katowice,
Poland}
\author{A.Budzanowski}
\affiliation{H. Niewodnicza{\'n}ski Institute of Nuclear Physics
PAN, Radzikowskiego 152, 31342 Krak{\'o}w, Poland}
\author{D.Filges}
\author{F.Goldenbaum}
\affiliation{Institut f{\"u}r Kernphysik, Forschungszentrum
J{\"u}lich, D-52425 J{\"u}lich, Germany}
\author{A.Heczko}
\affiliation{M. Smoluchowski Institute of Physics, Jagellonian
University, Reymonta 4, 30059 Krak{\'o}w, Poland}
\author{H.Hodde}
\affiliation{Institut f{\"ur} Strahlen- und Kernphysik, Bonn
University,  D-53121 Bonn, Germany}
\author{L.Jarczyk}
\author{B.Kamys }  \email{
ufkamys@cyf-kr.edu.pl [corresponding author]} \affiliation{M.
Smoluchowski Institute of Physics, Jagellonian University, Reymonta
4, 30059 Krak{\'o}w, Poland }
\author{M.Kistryn}
\affiliation{H. Niewodnicza{\'n}ski Institute of Nuclear Physics
PAN, Radzikowskiego 152, 31342 Krak{\'o}w, Poland}
\author{St.Kistryn}
\affiliation{M. Smoluchowski Institute of Physics, Jagellonian
University, Reymonta 4, 30059 Krak{\'o}w, Poland}
\author{St.Kliczewski}
\affiliation{H. Niewodnicza{\'n}ski Institute of Nuclear Physics
PAN, Radzikowskiego 152, 31342 Krak{\'o}w, Poland}
\author{A.Kowalczyk}
\affiliation{M. Smoluchowski Institute of Physics, Jagellonian
University, Reymonta 4, 30059 Krak{\'o}w, Poland}
\author{E.Kozik}
\affiliation{H. Niewodnicza{\'n}ski Institute of Nuclear Physics
PAN, Radzikowskiego 152, 31342 Krak{\'o}w, Poland}
\author{P.Kulessa}
\affiliation{Institut f{\"u}r Kernphysik, Forschungszentrum
J{\"u}lich, D-52425 J{\"u}lich, Germany} \affiliation{H.
Niewodnicza{\'n}ski Institute of Nuclear Physics PAN, Radzikowskiego
152, 31342 Krak{\'o}w, Poland}
\author{H.Machner}
\affiliation{Institut f{\"u}r Kernphysik, Forschungszentrum
J{\"u}lich, D-52425 J{\"u}lich, Germany}
\author{A.Magiera}
\author{W.Migda{\l}}
\affiliation{M. Smoluchowski Institute of Physics, Jagellonian
University, Reymonta 4, 30059 Krak{\'o}w, Poland}
\author{N.Paul}
\affiliation{Institut f{\"u}r Kernphysik, Forschungszentrum
J{\"u}lich, D-52425 J{\"u}lich, Germany}
\author{B.Piskor-Ignatowicz}
\affiliation{Institut f{\"u}r Kernphysik, Forschungszentrum
J{\"u}lich, D-52425 J{\"u}lich, Germany} \affiliation{M.
Smoluchowski Institute of Physics, Jagellonian University, Reymonta
4, 30059 Krak{\'o}w, Poland}
\author{M.Pucha{\l}a}
\affiliation{M. Smoluchowski Institute of Physics, Jagellonian
University, Reymonta 4, 30059 Krak{\'o}w, Poland}
\author{K.Pysz}
\affiliation{Institut f{\"u}r Kernphysik, Forschungszentrum
J{\"u}lich, D-52425 J{\"u}lich, Germany} \affiliation{H.
Niewodnicza{\'n}ski Institute of Nuclear Physics PAN, Radzikowskiego
152, 31342 Krak{\'o}w, Poland}
\author{Z.Rudy}
\affiliation{M. Smoluchowski Institute of Physics, Jagellonian
University, Reymonta 4, 30059 Krak{\'o}w, Poland}
\author{R.Siudak}
\affiliation{Institut f{\"u}r Kernphysik, Forschungszentrum
J{\"u}lich, D-52425 J{\"u}lich, Germany} \affiliation{H.
Niewodnicza{\'n}ski Institute of Nuclear Physics PAN, Radzikowskiego
152, 31342 Krak{\'o}w, Poland}
\author{M.Wojciechowski}
\affiliation{M. Smoluchowski Institute of Physics, Jagellonian
University, Reymonta 4, 30059 Krak{\'o}w, Poland}
\author{P. W{\"u}stner}
\affiliation{Institut f{\"u}r Kernphysik, Forschungszentrum
J{\"u}lich, D-52425 J{\"u}lich, Germany}

\collaboration{PISA - \textbf{P}roton \textbf{I}nduced
\textbf{S}p\textbf{A}llation collaboration}
%\homepage{http://www.kfa-juelich.de/ikp/pisa}

\date{\today}

\begin{abstract}
Energy and angular dependence of double differential cross sections
d$^2\sigma$/d$\Omega$dE was measured for reactions induced by 2.5
GeV protons on Au target with isotopic identification of light
products (H, He, Li, Be, and B) and with elemental identification of
heavier intermediate mass fragments (C, N, O, F, Ne, Na, Mg, and
Al).  It was found that two different reaction mechanisms give
comparable contributions to the cross sections. The intranuclear
cascade of nucleon-nucleon collisions followed by evaporation from
an equilibrated residuum describes low energy part of the energy
distributions whereas another reaction mechanism is responsible for
high energy part of the spectra of composite particles.
Phenomenological model description of the differential cross
sections by isotropic emission from two moving sources led to a very
good description of all measured data. Values of the extracted
parameters of the emitting sources are compatible with the
hypothesis claiming that the high energy particles emerge from
pre-equilibrium processes consisting in a breakup of the target into
three groups of nucleons; small, fast and hot fireball of $\sim$ 8
nucleons, and two larger, excited prefragments, which emits the
light charged particles and intermediate mass fragments. The smaller
of them contains $\sim$ 20 nucleons and moves with velocity larger
than the CM velocity of the proton projectile and the target. The
heavier prefragment behaves similarly as the heavy residuum of the
intranuclear cascade of nucleon-nucleon collisions.
%The mass and charge dependence of the total production cross
%sections was extracted from the above analysis for all observed
%reaction products. This dependence follows the power low behavior
%(A$^{-\tau}$ or Z$^{-\tau}$).
\end{abstract}

%    25.40-h  = Nucleon-induced reactions
%    25.40.Sc = Spallation
%    25.40.Ve = Other reactions above meson production threshold (400 MeV)


\maketitle

%-------------------------------------------------------------------

\section{\label{introduction} Introduction}

The mechanism of proton - nucleus interactions at GeV energies is
still not well understood.  Even the gold nucleus which is the most
frequently studied target, at least as concerns the measurements of
\emph{total cross sections} for emission of different products (cf.
Refs.\cite{KAU80A,ASA88A,CHE95A,WAD00A,REJ01A,BEN02A,BER02A,ROD02A,KAR03A} 
and references herein), reveals unexpected phenomena when more
exclusive observables are investigated. Recently measurements of
\emph{differential cross sections} in 4$\pi$ geometry were
undertaken \cite{HER06A,LET02A} for light charged particles (LCP's),
i.e. H and He ions, as well as for intermediate mass fragments
(IMF's) - Li and Be ions.  The measurements were done at 1.2 GeV
proton energy for $^{1,2,3}$H, $^{3,4,6}$He, $^{6,7,8,9}$Li and
$^{7,9,10}$Be isotopes \cite{HER06A}, and at 2.5 GeV for
$^{1,2,3}$H, $^{3,4}$He, and $^{6,7}$Li isotopes \cite{LET02A}.

It was found that the shape of energy spectra of emitted composite
 particles as well as their angular dependence cannot
be explained using the conventional picture of the intranuclear
cascade of nucleon-nucleon collisions followed by fragment
evaporation from excited remnant nucleus in competition with fission
process. Whereas the low energy part of spectra - up to 60 - 80 MeV
- seems to be reasonably well described by this conventional
mechanism, the high energy part of spectra is generally strongly
underestimated by any of the existing models.
%behaves in quite another manner.
It is more flat than the low energy part of the spectrum and its
slope increases monotonically with the emission angle. This behavior
indicates the necessity to include non-equilibrium processes in the
description of the reaction mechanism. Authors of Refs.
\cite{LET02A,BOU04A,HER06A} propose the surface coalescence of
emitted nucleons as process responsible for high energy part of the
$^{2,3}$H and $^3$He spectra. They claim, however, that such a
mechanism is ruled out for $^4$He and heavier ejectiles
\cite{LET02A,HER06A}.

      In the present study the task was undertaken
to measure double differential cross sections $(d^2\sigma/d\Omega
dE)$ with isotopic identification of the light reaction products
from proton - gold collisions at proton beam energy of 2.5 GeV,
extending the range of detected ejectiles to heavier than those from
previous reported investigations \cite{LET02A,HER06A}.

 It should be emphasized that for the gold target the double
differential cross sections $(d^2\sigma/d\Omega dE)$  of
intermediate mass fragments, i.e. fragments with mass number $A_F
>4$, were not measured up to now with isotopic identification . The
only available data are published by Letourneau et al. \cite{LET02A}
for $^{6,7}$Li. Low statistics of isotopically identified data in
publication of Herbach et al. \cite{HER06A}  did not allow to
analyze double differential cross sections $(d^2\sigma/d\Omega dE)$
for isotopes heavier than $^4$He. For individual isotopes only the
analysis of angle integrated $(d\sigma/dE)$ spectra or energy
integrated $(d\sigma/d\Omega)$ angular distributions was possible.

The goal of the present study was to gain new experimental
information on the proton - gold interaction at proton energy of 2.5
GeV. Those new double differential data $(d^2\sigma/d\Omega dE)$
should allow to gain deeper insight in the mechanism of
non-equilibrium processes.
%Thus, in the present study the energy spectra
%were measured by seven telescopes built of silicon semiconductor
%detectors placed at 16, 20, 35, 50, 65, 80, and 100 deg.  Some of
%them (those at angles 16, 20, 65 \fbox{and ???} ) were accompanied
%by CsI scintillator detector 7 cm thick, what enabled us to measure
%broad range of energies for LCP's.
%
%\begin{center}
%\fbox{ (??? Bragg detectors ???)}
%\end{center}

Details of the experimental procedure are discussed in the second
section and the obtained data in the next, third section.  The
fourth section is devoted to model description of the measured
spectra. The interpretation, summary and conclusions are presented
in the last section. Formulae applied in the phenomenological
parametrization are collected in the appendix.

\section{\label{experiment} Experimental procedure}

     The experiment has been performed using the internal beam of the Cooler
Synchrotron COSY in the Research Center in J\"ulich. Due to multiple
passing of the circulating internal beam through the target it was
possible to achieve as high luminosity as that which can be reached
only with very intensive external beam of accelerators (with the
particle current of order of mA).  Circulation of the beam without
its immediate absorption demanded using of very thin, self
supporting targets (of order of 300 $\mu$g/cm$^2$) what in turn
resulted in negligible distortion of the reaction product spectra by
interaction of the emitted particles with the target. Furthermore,
during each cycle of injection and acceleration, the protons were
circulating in the COSY ring slightly below the target, being slowly
bumped onto the target until the beam was completely used up. Then a
new cycle was started. The speed of the vertical shift of the proton
beam was controlled by feedback of the observed reaction rate to
avoid overloading of the data acquisition system.

The scattering chamber and the detecting system was described in
detail in ref. \cite{BAR04A}. There, however, the main interest was
focused only on performance of the gridded ionization chambers which
are used in the experiment for charge identification as well as for
energy measurement of the reaction products by means of the Bragg
curve spectroscopy. For this reason, a description of the other
components of the detection system, relevant to the data discussed
in the present paper, will be given here in a more detailed way.

The PISA apparatus consists of nine independent detection arms
comprising various kinds of detectors. Two of these arms (placed at
15 and 120$^{\circ}$ angles in respect to the beam direction) are
equipped with the Bragg curve detectors (BCD), which
%together with
%the time-of-flight measuring systems (telescopes of two
%micro-channel plate detectors) positioned in front of the BCD's
%entrance window,
permit the \emph{Z}-identification of the reaction products and
determination of their kinetic energies with low detection energy
threshold (of about 1 MeV/nucleon).
%Telescopes of
%silicon detectors installed behind the active volume of the BCD,
%together \underline{with the following them scintillation detectors
%(?)} increase the detection energy range of these two detection arms
%for the light and fast spallation product, namely for $H$ and $He$
%isotopes (as well as for the lightest intermediate mass fragments,
%i.e. $Li$ nuclei).  Heavier fragments are stopped in the gas of BCD
%or in the silicon detectors.\\
%
The telescopes consisted of silicon detectors are installed at the
detection angles of 35, 50, 80 and 100$^{\circ}$. The detectors
operate in the ultra high vacuum (UHV) of the COSY accelerator and
are cooled to a temperature of -10$^{\circ}$ C. The cooling improves
the energy resolution of the detectors, thus the isotopes of all
ejectiles up to carbon can be identified. Due to geometrical
constraints the silicon telescope detectors placed in the vacuum at
35, 50, and 80$^{\circ}$ cannot be supplied with additional
detectors. Consequently light charged particles of high energies,
not being stopped in the silicon detectors, cannot be detected.
%do not allow to supplement them with additional detectors, thus they
%cannot be used in a full analysis of light charged particles of high
%energies which are not stopped in the silicon detectors.
The upper limit of energies of particles stopped by these telescopes
is about 30 - 40 MeV for protons, deuterons and tritons but
increases significantly for heavier particles, e.g. for alpha
particles it is around 120 MeV. Therefore these telescopes are
suitable to measure the low energy part of the spectra for hydrogen
isotopes, a large range of energies for helium isotopes and the full
energy spectra of intermediate mass fragments. The silicon detector
telescope placed at 100$^{\circ}$ in the ultra high vacuum has
another construction than the telescopes mentioned above, thus it
was possible to supplement it with a 7.5 cm thick CsI scintillator
detector standing behind it in the air (outside the chamber),
separated by a steel window of 50 $\mu$m thickness from the ultra
high vacuum of COSY.
At three angles (15.6, 20, and 65$^{\circ}$), the telescopes built
of silicon detectors with 7.5 cm CsI scintillator detectors standing
behind them are positioned outside the chamber.  The destination of
these telescopes as well as that at 100$^{\circ}$ is to
significantly increase range of energies of detected light charged
particles and IMF's.

\begin{table*}
%  \centering
  \caption{\label{table:energies_az}Range of energies (in MeV) of isotopically identified
  reaction products detected at various scattering angles }
\begin{tabular}[b]{|l||c|c|c|c|c|c|c|}
  \hline
    & \multicolumn{7}{c|}{Angle [degrees]}   \\
  \cline{2-8}
 Ejectile   &   15.6                &   20          &   35          &   50          &    65         &   80          &   100 \\
  \hline
  \hline
  \emph{p}         &   7.5 -- 103.5 &  7.5 -- 92.5  &  3.5 -- 21.5  &  3.5 -- 21.5  &  7.5 -- 97.5  &  3.5  -- 6.5  &  9.5 -- 120.5 \\
  \hline
  \emph{d }        &   8.5 -- 210.5 &  8.5 -- 210.5 &  5.5 -- 35.5  &  6.5 -- 35.5  &  8.5 -- 212.5 &  5.5  -- 8.5  & 12.5 -- 218.5  \\
  \hline
  \emph{t }        &   9.5 -- 240.5 & 11.5 -- 242.5 &  4.5 -- 37.5  &  6.5 -- 33.5  &  9.5 -- 249.5 &  4.5  -- 10.5 & 14.5 -- 162.5 \\
  \hline
  $^3$He    &  20.5 -- 296.5 & 20.5 -- 296.5 &  8.5 -- 95.5  &  8.5 -- 86.5  & 20.5 -- 292.0 &  13.5 -- 19.5 &  9.5 -- 161.5 \\
  \hline
  $^4$He    &  23.5 -- 277.5 & 23.5 -- 253.0 &  9.5 -- 120.5 &  8.5 -- 113.5 & 23.5 -- 182.5 &  13.5 -- 25.5 &  9.5 -- 122.5 \\
  \hline
  $^6$He    &  26.5 -- 83.5  & 26.5 --  74.5 & 11.5 -- 122.5 & 10.5 -- 106.5 & 26.5 --  77.5 &  15.5 -- 24.5 & 11.5 --  53.5 \\
  \hline
  $^6$Li    &  43.5 -- 145.5 & 42.5 -- 147.5 & 17.5 -- 178.0 & 14.5 -- 178.0 & 41.5 -- 143.5 &  18.5 -- 48.5 & 18.5 -- 105.5 \\
  \hline
  $^7$Li    &  44.5 -- 155.5 & 35.5 -- 156.5 & 18.5 -- 159.5 & 16.5 -- 136.5 & 44.5 -- 152.5 &  19.5 -- 55.5 & 18.5 -- 117.5 \\
  \hline
  $^8$Li    &  47.5 -- 113.5 & 47.5 -- 110.5 & 19.5 -- 115.5 & 17.5 -- 98.5  & 46.5 -- 112.5 &  21.5 -- 51.5 & 19.5 -- 85.5 \\
  \hline
  $^9$Li    &  49.5 -- 85.5  & 49.5 -- 118.5 & 19.5 -- 82.5  & 17.5 -- 53.5  & 49.5 -- 85.5  &  22.5 -- 52.5 & 20.5 -- 65.5 \\
  \hline
  $^7$Be    &  61.5 -- 136.5 & 62.5 -- 146.5 & 24.5 -- 123.5 & 24.5 -- 138.5 & 61.5 -- 136.5 &  27.5 -- 69.5 & 27.5 -- 90.5 \\
  \hline
  $^9$Be    &  68.5 -- 116.5 & 68.5 -- 119.5 & 25.5 -- 94.5  & 25.5 -- 94.5  & 68.5 -- 107.5 &  29.5 -- 80.5 & 27.5 -- 84.5 \\
  \hline
  $^{10}$Be &  71.5 -- 116.5 & 71.5 -- 128.5 & 26.5 -- 101.5 & 23.5 -- 98.5  & 71.5 -- 122.5 &  30.5 -- 87.5 & 29.5 -- 80.5 \\
  \hline
  $^{10}$B  &  90.5 -- 123.5 & 92.5 -- 122.5 & 35.5 -- 92.5  & 30.5 -- 99.5  & 90.5 -- 111.5 &  38.5 -- 86.5 & 36.5 -- 90.5 \\
  \hline
  $^{11}$B  &  94.5 -- 136.5 & 94.5 -- 130.5 & 35.5 -- 116.5 & 31.5 -- 100.5 & 96.5 -- 114.5 & 39.5 -- 105.5 & 37.5 -- 91.5 \\
  \hline
  $^{12}$B  &                &               & 36.5 -- 96.5  & 35.5 --83.5   &               & 41.5 -- 83.5  & 39.5 -- 78.5 \\
  \hline

\end{tabular}
\end{table*}

\begin{table*}
%  \centering
  \caption{\label{table:energies_z}Range of energies (in MeV) of elementally identified
  reaction products detected at various scattering angles }
\begin{tabular}[b]{|l||c|c|c|c|c|c|c|}
  \hline
    & \multicolumn{6}{c|}{Angle [degrees]}   \\
  \cline{2-7}
 Ejectile   &   15          &   35          &  50           &   80          &   100        &  120 \\
  \hline
  \hline
  C         & 11.5 -- 136.5 & 46.5 -- 118.5 & 40.5 -- 118.5 & 50.5 -- 116.5 & 48.5 -- 102.5 & 12.5 -- 57.5 \\
  \hline
  N         & 14.5 -- 69.5  & 56.5 -- 116.5 & 48.5 -- 108.5 & 61.5 -- 109.5 & 57.5 -- 99.5  & 15.5 -- 70.5 \\
  \hline
  O         & 15.5 -- 80.5  & 67.5 -- 103.5 & 58.5 -- 103.5 & 74.5 -- 119.5 & 69.5 -- 99.5  & 18.5 -- 68.5 \\
  \hline
  F         & 21.5 -- 98.5  &               &               &               &               & 22.5 -- 88.5 \\
  \hline
  Ne        & 25.5 -- 109.5 &               &               &               &               & 23.5 -- 100.5 \\
  \hline
  Na        & 29.5 -- 127.5 &               &               &               &               & 26.5 -- 117.5 \\
  \hline
  Mg        & 29.5 -- 106.5 &               &               &               &               & 28.5 -- 98.5 \\
  \hline
  Al        & 31.5 -- 94.5  &               &               &               &               & 30.5 -- 72.5 \\
  \hline

\end{tabular}
\end{table*}

%\begin{figure}
%\begin{center}
%\includegraphics[angle=0,width=0.5\textwidth]{FIGURES/be7norm.eps}
%\caption{Examples of the energy spectra of $^7$Be measured in the
%present experiment (open symbols) for 16, 35, 65, and 100 deg. The
%lines show phenomenological parametrization described in the next
%section of the paper.}\label{fig:be7norm}
%\end{center}
%\end{figure}

The particles observed at different angles in the present experiment
and the energy ranges covered by the detecting system are listed in
the Table \ref{table:energies_az}  and in the Table
\ref{table:energies_z} for isotopically identified and elementally
identified ejectiles, respectively.

The absolute normalization of the data was achieved by comparison of
the total cross section for $^7$Be ejectiles extracted from angular
and energy integration of the spectra measured in the present
experiment with the cross section obtained from parametrization of
experimental $^7$Be production cross sections in proton-nucleus
collisions, ref. \cite{BUB04A}.
%
%The integration of spectra was performed in two steps; first, the
%phenomenological formula of single moving source was applied to
%description of the data and then this analytic formula with fixed
%values of the parameters from the above fit was used to get the
%total cross section. The quality of parametrization of data is
%illustrated by Fig. \ref{fig:be7norm}.
%
Accuracy of the absolute normalization was estimated to be better
than 10\%.

\section{\label{results} Experimental results}

\begin{figure}
\begin{center}
\includegraphics[angle=0,width=0.52\textwidth]{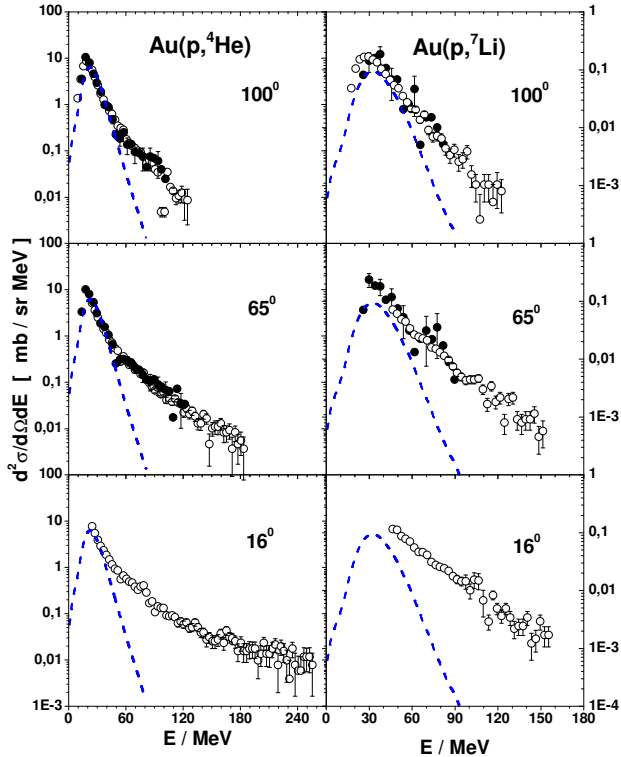}
\caption{\label{fig:he4li7buu} Typical energy spectra of $^4$He
(left column) and $^7$Li particles (right column) measured in the
present experiment (open circles) and published in Ref.
\cite{LET02A} (full dots) for corresponding emission angles. The
lines show prediction of evaporation of $^4$He and $^7$Li evaluated
by means of GEM program of S. Furihata \cite{FUR00A,FUR02A} from
excited residual nuclei of the first stage of the reaction with
properties extracted from BUU calculations. }
\end{center}
\end{figure}

In the present study the double differential spectra
$d^2\sigma/d\Omega dE$ were determined for the first time for many
isotopically identified intermediate mass fragments emitted from
proton - gold collisions at GeV energies. This concerns $^6$He,
$^{8,9}$Li, $^{7,9,10}$Be, $^{10,11,12}$B spectra which were not
measured by Letourneau \emph{et al.} at 2.5 GeV \cite{LET02A}
whereas the experiment of Herbach \emph{et al.} \cite{HER06A} at 1.2
GeV which detected IMF's lighter than boron had statistics allowing
to extract only elemental spectra. Typical spectra for isotopically
identified particles from the present experiment are shown in Fig.
\ref{fig:he4li7buu} together with data measured by Letourneau
\emph{et al.} \cite{LET02A}. Excellent agreement of the present data
with those published by Letourneau \emph{et al.} was achieved for
all products measured in both experiments, i.e. $^{1,2,3}$H,
$^{3,4}$He, and $^{6,7}$Li. Note, that statistical errors, which are
only shown for selected $^7$Li data of Ref. \cite{LET02A}, present
indeed typical errors for all $^7$Li data from that paper.

The energy distributions of emitted ejectiles have shapes resembling
Maxwellian evaporation spectra, but because of an instrumental
low-energy cutoff it was not possible to observe the maxima of these
distributions for fragments heavier than boron. Although for heavier
fragments the Coulomb barrier moves the position of maximum  of the
yield towards higher values, the large energy loss in first silicon
detector of telescope prevent us from detecting heavy ejectiles at
energies close to the maximum of the energy distributions. To avoid
this problem, i.e. to measure low energy part of the spectra, two
Bragg curve ionization chambers (BCD) were applied. They were placed
at 15$^0$ and 120$^0$ scattering angles. Since BCD's allow for the
identification of the charge of ejectiles only, the spectra of heavy
products, i.e. C, O, N, F, Ne, Na, Mg and Al were obtained only with
elemental identification. As far as we know, similar spectra were up
to now investigated for the gold target only in the experiment
performed at 1.0 GeV by Kotov \emph{et al.} \cite{KOT95A}, where
double differential cross sections were measured without isotopic
identification.

Typical properties of the spectra, characteristic for all ejectiles,
are depicted in Fig. \ref{fig:he4li7buu}. At low energy the angle
independent - Maxwellian like - contribution is well visible. This
isotropic energy distribution  may be reproduced by the two stage
model discussed below.  The dashed line shown in the figure,
represents predictions of this model.  Another contribution, i.e. an
exponential distribution, strongly varying with angle is present at
higher energy in all experimental spectra. The slope of this
anisotropic energy contribution increases with the angle, what may
be interpreted as effect of fast motion of
an emitting source in the forward direction.\\

\section{\label{analysis} Theoretical analysis}

In the most frequently considered scenario of the proton-nucleus
collision at GeV proton energies it is assumed that reaction
proceeds via two stages.

In the first stage of the reaction the proton impinging on to the
target nucleus initiates a cascade of nucleon-nucleon collisions
which leads to emission of several fast nucleons and pions, and to
excitation of the nucleus. This fast stage of reaction is described
by intranuclear cascade (INC) model, e.g. \cite{CUG97A,BOU02A},
Boltzmann-Uehling-Uhlenbeck (BUU) model, e.g. \cite{BER88A} or by
quantum molecular dynamics (QMD) model, e.g. \cite{AIC91A}. The
first of the mentioned models gives an account of the
nucleon-nucleus interaction by static (time-independent) mean field,
the BUU allows for dynamic evolution of the mean field as caused by
time dependence of an average nucleon density, and the QMD treats
the nucleon-nucleus interaction as a time dependent sum of
elementary two-nucleon and three-nucleon interactions of all
nucleons. The QMD introduces the largest fluctuations of the density
distribution of nucleons  and, therefore, allows for emission of
clusters of nucleons from the first stage of the reaction. The
static mean field description used by INC model automatically
precludes possibility of nucleon distribution fluctuations. The BUU
model takes into consideration a time dependent modification of the
nucleon density distribution, however, the averaging over many test
particles, present inherently in BUU, prohibits appearing of
fluctuations large enough for nucleon clusters emission. The
emission of fast nucleons (in the case of INC and BUU) or fast
nucleons and light clusters (in the case of QMD) terminates after a
short time, leaving the residual excited nucleus in a status close
to the thermodynamic equilibrium.

\begin{figure}
\begin{center}
\includegraphics[angle=0,width=0.5\textwidth]{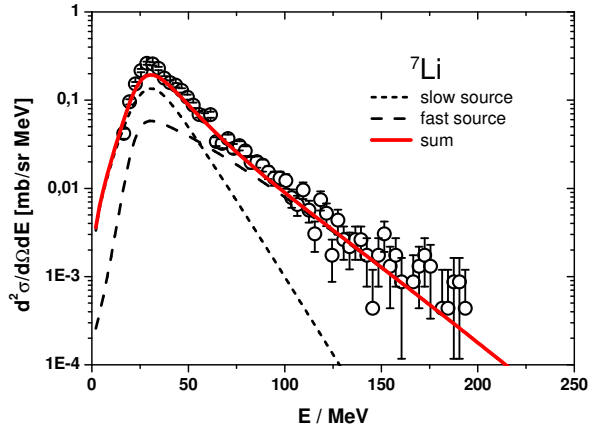}
\caption{\label{fig:twosrcs} Open circles represent experimental
energy spectrum of $^7Li$ particles measured at 35$^0$ in the
current experiment.  The lines present result of phenomenological
model described below; the short-dashed line shows contribution of
the slow emitting source, the long-dashed line depicts contribution
of the fast source, whereas the solid line presents sum of both
contributions. Please note, that the shape of this experimental
energy distribution as compared with spectra shown in right part of
Fig. \ref{fig:he4li7buu} also confirms the monotonic change of the
exponential slope with the scattering angle.}
\end{center}
\end{figure}

\begin{figure}
\begin{center}
\includegraphics[angle=0,width=0.5\textwidth]{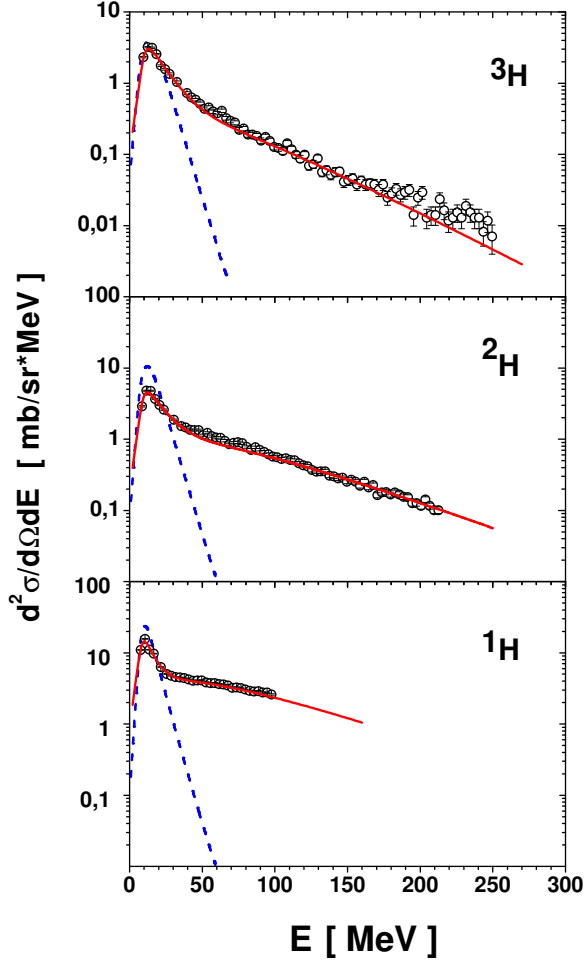}
\caption{\label{fig:h1232s} Open circles represent typical spectra
of protons, deuterons and tritons measured in the present experiment
by telescope consisted of silicon semiconductor detectors and 7.5 cm
thick scintillating detector CsI  placed at scattering angle of 65
degree in respect to the proton beam. The dashed lines show
evaporation contribution evaluated by means of the BUU and
Generalized Evaporation Model whereas the full lines correspond to
phenomenological model of two emitting sources described below. Note
change of the scale for the triton spectrum.}
\end{center}
\end{figure}

The second stage of reaction consists in the evaporation of nucleons
and clusters from this equilibrated system, which can also undergo
fission with emission of two heavy fragments. Thus, in the two-step
model of reaction mechanism, the non equilibrium emission of
nucleons and clusters can appear only in the first stage of the
reaction. It is believed that statistical model codes like, e.g. GEM
\cite{FUR00A,FUR02A} or GEMINI \cite{CHA88A} are capable to well
reproduce emission of nucleons and fragments from equilibrated,
excited nucleus. Therefore, observation of any disagreement of the
data with predictions of the two-step model would suggest that (i)
the model is not adequate to the real situation (e.g. an additional,
intermediate stage of the process is necessary before achieving
thermodynamic equilibrium), or (ii) description of the emission of
particles from the first stage of the reaction (nucleons or
clusters)
is not properly taken into consideration.\\

\begin{figure}
\begin{center}
\includegraphics[angle=0,width=0.5\textwidth]{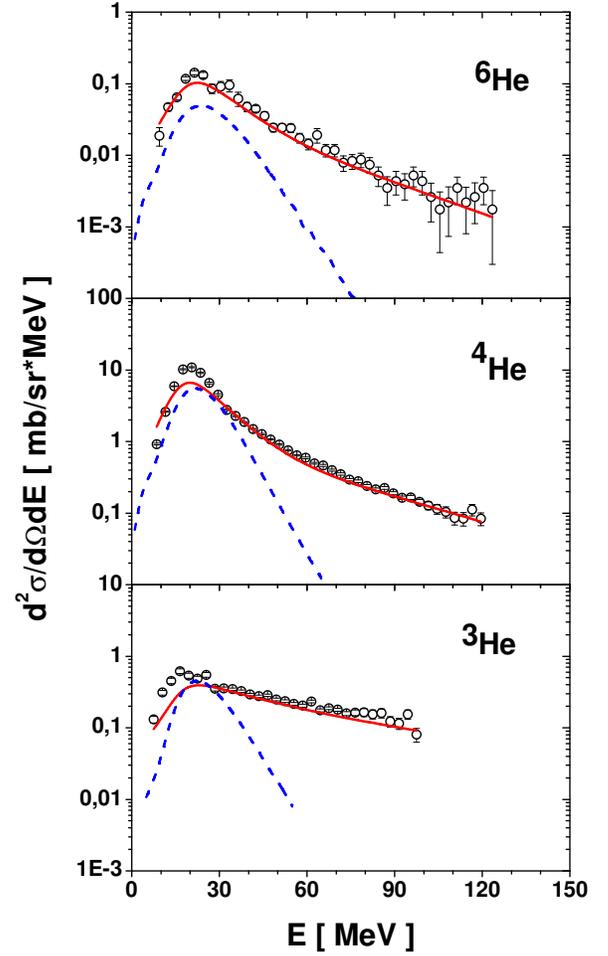}
\caption{\label{fig:he346} Typical energy spectra of helium ions
$^{3,4,6}$He measured in the present experiment by telescope
consisted of silicon semiconductor detectors placed at scattering
angle of 35 degree in respect to the proton beam - open circles.
Note different vertical scales for each spectrum. The lines have the
same meaning as in Fig. \ref{fig:h1232s}}
\end{center}
\end{figure}

\subsection{Boltzmann-Uehling-Uhlenbeck model and evaporation model}

\begin{figure}
\begin{center}
\includegraphics[angle=0,width=0.5\textwidth]{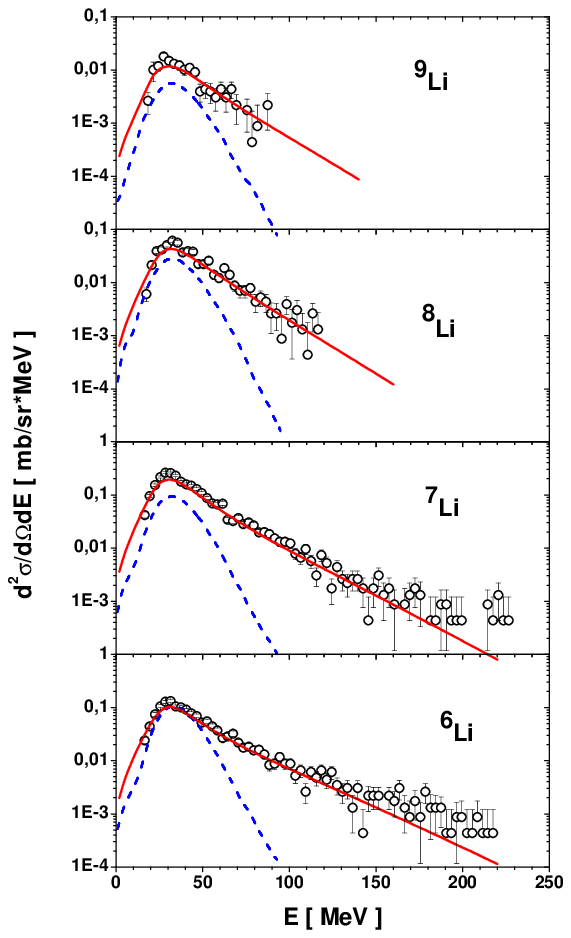}
\caption{\label{fig:li6789_2s035} Typical spectra of lithium ions
$^{6,7,8,9}$Li measured in the present experiment by telescope
consisted of silicon semiconductor detectors  placed at scattering
angle of 35 degree in respect to the proton beam - open circles.
Note different vertical scales for $^{6,7}$Li and $^{8,9}$Li. The
lines have the same meaning as in Fig. \ref{fig:h1232s}}
\end{center}
\end{figure}

\begin{figure}
\begin{center}
\includegraphics[angle=0,width=0.5\textwidth]{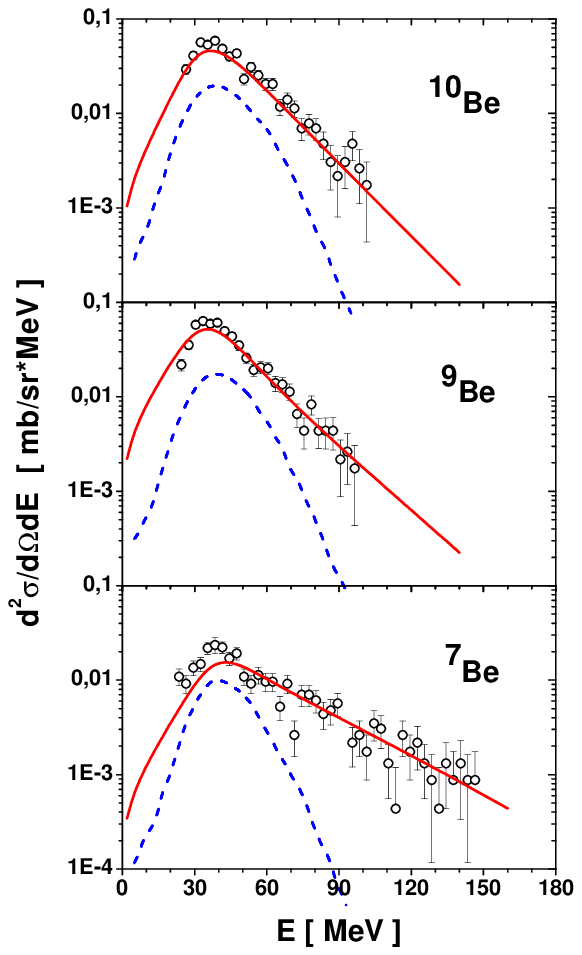}
\caption{\label{fig:be7910_2s035} Typical spectra of beryllium ions
$^{7,9,10}$Be measured in the present experiment by telescope
consisted of silicon semiconductor detectors  placed at scattering
angle of 35 degree in respect to the proton beam - open circles. The
lines have the same meaning as in Fig. \ref{fig:h1232s}}
\end{center}
\end{figure}

\begin{figure}
\begin{center}
\includegraphics[angle=0,width=0.5\textwidth]{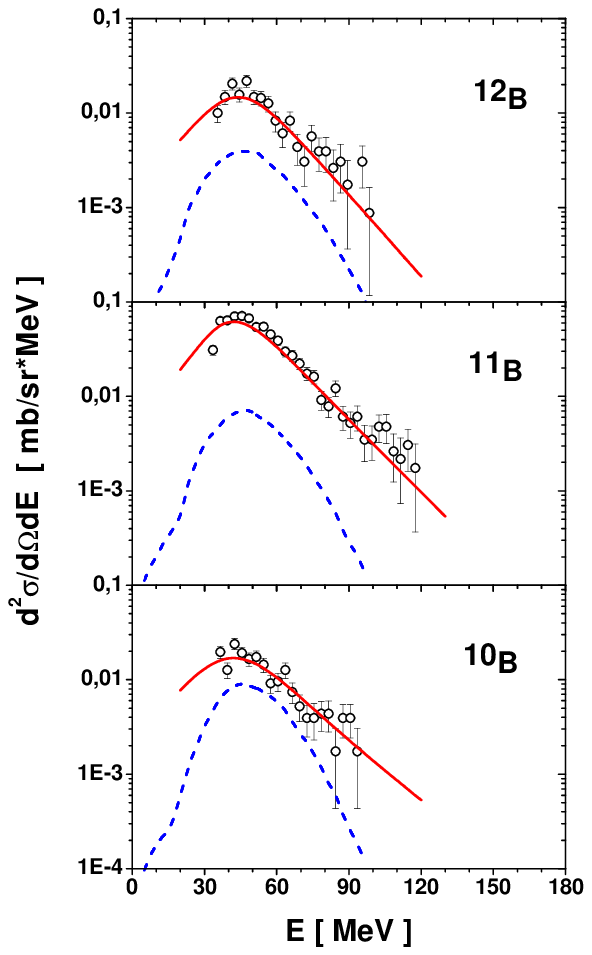}
\caption{\label{fig:b10_12_2s035} Typical spectra of boron ions
$^{10,11,12}$B measured in the present experiment by telescope
consisted of silicon semiconductor detectors  placed at scattering
angle of 35 degree in respect to the proton beam - open circles. The
lines have the same meaning as in Fig. \ref{fig:h1232s}}
\end{center}
\end{figure}

\begin{figure}
\begin{center}
\includegraphics[angle=0,width=0.5\textwidth]{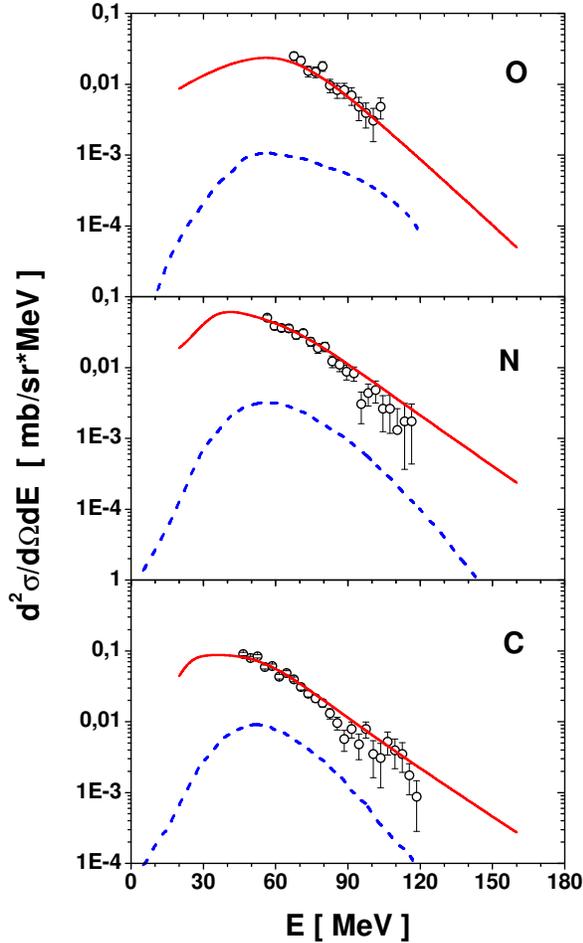}
\caption{\label{fig:cno_2s035} Typical spectra of carbon, nitrogen
and oxygen ions measured  in the present experiment without isotopic
separation by telescope consisted of silicon semiconductor detectors
placed at scattering angle of 35 degree in respect to the proton
beam - open circles.  The lines have the same meaning as in Fig.
\ref{fig:h1232s}}
\end{center}
\end{figure}

The present data were compared with results of a two stage model in
which the Boltzmann-Uehling-Uhlenbeck transport equation
\cite{BER88A} has been applied for the description of the first step
of the proton - nucleus collision leading to emission of fast
nucleons leaving the heavy excited remnant in a state close to
equilibrium. Monte Carlo computer program developed at Giessen
University \cite{CAS06A} was utilized to simulate this first stage
of the reaction and to find properties of excited residual nuclei.
Deexcitation of these nuclei, which proceeds by emission of nucleons
and complex fragments, was calculated in the frame of statistical
model using the GEM (Generalized Evaporation Model) computer program
of Furihata \cite{FUR00A,FUR02A}.  Theoretical energy spectra of
various ejectiles found from this two stage model are shown in Figs.
\ref{fig:he4li7buu} - \ref{fig:cno_2s035} as dashed lines.  It can
be concluded from examination of these figures that the model
predictions describe well low energy part of spectra for hydrogen,
helium and lithium isotopes.  For heavier ejectiles the theoretical
cross sections underestimate the experimental data. Moreover, it can
be observed that the high energy part of the spectra is clearly not
reproduced by the two stage model, which predicts much steeper slope
of the spectra than is observed experimentally. Thus, another
mechanism seems to give a significant contribution to the proton -
nucleus reactions.
As concerns hydrogen and helium production, the authors of
\cite{HER06A}, \cite{LET02A}, and \cite{BOU04A} papers propose the
coalescence of nucleons as the mechanism responsible for this
effect, however, no microscopic model is able to reproduce observed
effects for heavier composite ejectiles.

An extensive comparison of predictions resulting from the models
mentioned above with our experimental data presented here will be
subject of a forthcoming paper. We restrict ourselves here on
conclusions we can draw from the application of a phenomenological
model described in the next section.

Following properties of the spectra should be taken into
consideration when looking for an appropriate phenomenological
model:
\begin{description}
  \item[(i)] \hspace{0.5em} The position
   of the peak present at low energies
   in the experimental spectra of all observed particles (and its height
   for light ejectiles) is quite well reproduced by the two stage model discussed above.
   This means that the mechanism described by this model gives a large
   contribution to the reaction and therefore it must be
   taken into account in the frame of any phenomenological model.
  \item[(ii)] \hspace*{0.4em} The slope of the exponential, high energy tail
  of the
   experimental spectra for all ejectiles varies monotonically, increasing
   strongly with the scattering angle
   as can be seen from Fig. \ref{fig:he4li7buu}.  Such a behavior is
   in contradistinction to properties of the spectra evaluated in
   the frame of the two stage model, which are almost independent of angle.
   This indicates that high energy particles are not emitted from heavy residuum
   of the intranuclear cascade but from another source which moves much
   faster than the residuum.
\end{description}

These arguments call for using of a phenomenological model of two
emitting sources; one source moving slowly would imitate emission
from heavy residuum of the intranuclear cascade whereas the second
source should simulate emission from faster (and thus probably
lighter) nuclear system. Of course, one could imagine that more than
two sources of emitted particles are necessary for reasonable
description of the data. The applied model of two moving sources
corresponds to minimal number of parameters necessary to fulfill
qualitative demands put on the model by the experimental data.

\begin{figure}
  \includegraphics[width=0.48\textwidth]{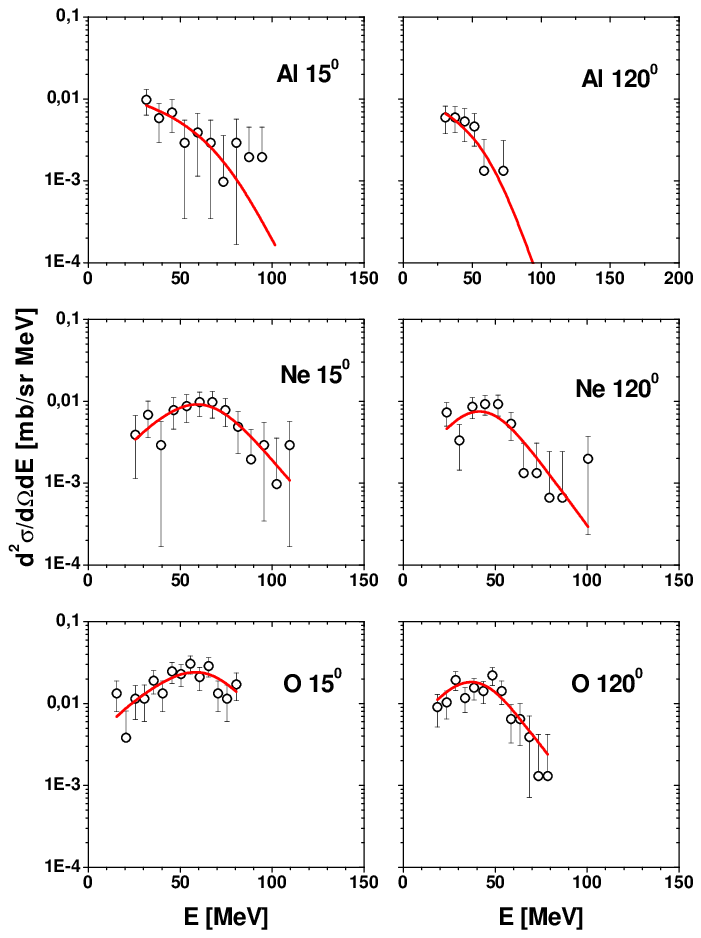}\\
  \caption{Examples of energy spectra for heavier, elementally identified IMF's
  obtained in the current experiment by means of Bragg curve ionization
  chambers.  The open circles represent experimental data and the solid lines
  show results of phenomenological model analysis obtained with assumption
  of only one moving source, discussed in the next
  section of the paper.}\label{fig:bcdfig.eps}
\end{figure}

\subsection{Phenomenological model of two moving sources}

In the frame of the phenomenological model of two moving sources the
angular and energy dependence of the double differential cross
sections $d^2 \sigma/ d\Omega dE$ is described by analytical
formulae. The details of the model and interpretation of its
parameters are presented in the Appendix. An example of the
description of the experimental energy spectrum by the two source
model is shown in Fig. \ref{fig:twosrcs}. The symbols depict the
data from present experiment obtained for $^7$Li ejectiles detected
at scattering angle 35$^0$ whereas the lines show result of the fit
of the phenomenological model. The short-dashed line presents
contribution from the slowly moving source, the long-dashed line
shows contribution from the fast source and the solid line
corresponds to sum of both contributions. As can be seen, very good
description of full energy spectrum could be achieved.

The parameters of the theoretical formula of the two moving sources
model have been searched for by fitting simultaneously
%, by the same set of parameters,
experimental spectra at several scattering angles  for each
ejectile. Exceptions from this rule were the spectra of ejectiles
heavier than oxygen (F, Ne, Na, Mg, and Al), which were measured
only at these two angles at which Bragg curve ionization chambers
have been positioned, \emph{i.e}. at 15$^0$ and 120$^0$.  Such
spectra were fitted assuming that only one moving source gives
contribution to the reaction.
Furthermore, the spectra of C, N, and O which were measured both, by
silicon detectors at 35$^0$, 50$^0$, 80$^0$, and 100$^0$ as well as
by Bragg curve ionization chambers at 15$^0$ and 120$^0$ degree were
fitted using one emitting source and two emitting sources.  The
parameters of sources for light charged particles and isotopically
identified IMF's  are listed in Table \ref{table:parameters} whereas
those for heavier IMF's, which were only elementally identified, are
collected in  Table \ref{table:elements}.

The first source should simulate evaporation of particles  from
heavy remnant of the first stage of the reaction, \emph{i.e.}
intranuclear cascade of nucleon-nucleon collisions.  Thus, its
velocity was fixed at value $\beta$=0.002 (in units of velocity of
light) as it was extracted from BUU calculations.  This value was
constant for all calculations. Other parameters characterizing the
source, \emph{i.e.} \emph{k}-parameter (ratio of the actual height
of Coulomb barrier to its value found from simple estimation for two
touching, charged spheres), \emph{T}-parameter (apparent
temperature), and $\sigma$ (energy and angle integrated cross
section for production of given ejectile) were free parameters of
the fit.

All parameters of the second source were freely modified in fits
since no hypothesis concerning origin of this source was made before
the analysis. Usually the program was able to find unambiguously the
best parameters, corresponding to the minimum value of chi-square.
In such a situation the routine provides estimation of errors.  In
some cases, however, the valley of chi-square values was so
complicated that the program was not able to produce reasonable
estimation of errors. The ambiguity of parameters lead sometimes the
searching procedure to nonphysical values of the parameters, as
\emph{e.g.} negative height of the Coulomb barrier.  Then it was
necessary to fix these parameters at values, which still have
physical meaning. Such values of parameters are quoted in the tables
as closed in square parentheses.

% rotated 90 deg table - NOTE: "rotating" package is necessary for this purpose
%=====================----------------------------------------------------------
%
%\begin{sidewaystable}
\begin{table*}
%  \centering
  \caption{\label{table:parameters}Parameters of two moving sources for isotopically identified products }
\begin{tabular}{|l||l|l|l||l|l|l|l||c|}
  \hline
            & \multicolumn{3}{c||}{Slow source} & \multicolumn{4}{c||}{Fast source}                 &  \\
  \cline{2-8}
  Ejectile  & \emph{k}$_1$    & \emph{T}$_1$/MeV & $\sigma_1$/mb & \emph{k}$_2$  &     $\beta_2$   & \emph{T}$_2$/MeV         & $\sigma_2$/mb & $\chi^2$ \\
  \hline
  \hline
  \emph{p}  & 0.67 $\pm$ 0.02 & 5.6  $\pm$ 0.3 & 1712 $\pm$ 46  & [0.05]          & 0.147 $\pm$ 0.005   & 51.0 $\pm$ 1.4 & 4839 $\pm$ 86  & 22.6 \\
  \hline
  \emph{d}  & 0.75 $\pm$ 0.02 & 9.2  $\pm$ 0.4 & 870  $\pm$ 29  & 0.07 $\pm$ 0.01 & 0.127 $\pm$ 0.004   & 48.3 $\pm$ 0.8 & 1100 $\pm$ 24  & 13.2 \\
  \hline
  \emph{t}  & 0.85 $\pm$ 0.02 & 9.5  $\pm$ 0.3 & 627  $\pm$ 17  & [0.05]          & 0.072 $\pm$ 0.003   & 36.1 $\pm$ 0.7 &  323 $\pm$ 13  &  6.1 \\
%  \hline
  \hline
  $^3$He    & 0.75 $\pm$ 0.03 & 14.9 $\pm$ 0.8 & 112  $\pm$  7  & [0.05]          & 0.083 $\pm$ 0.005   & 45.5 $\pm$ 1.2 &  106 $\pm$  7  &  4.6 \\
  \hline
  $^4$He    & 0.82 $\pm$ 0.02 &  7.8 $\pm$ 0.3 & 1722 $\pm$ 43  & 0.30 $\pm$ 0.09 & 0.048 $\pm$ 0.005   & 26.6 $\pm$ 1.5 &  251 $\pm$ 30  & 59 \\
  \hline
  $^6$He    & 0.97 $\pm$ 0.04 &  9.0 $\pm$ 0.6 & 24.8 $\pm$ 1.4 & 0.35 $\pm$ 0.05 & 0.040 $\pm$ 0.007   & 21.6 $\pm$ 1.4 &  7.5 $\pm$ 1.4 & 2.1 \\
%  \hline
  \hline
  $^6$Li    & 0.86 $\pm$ 0.04 & 11.1 $\pm$ 0.8 & 25.3 $\pm$ 1.7 & 0.44 $\pm$ 0.04 & 0.034 $\pm$ 0.003   & 23.7 $\pm$ 0.6 & 14.5 $\pm$ 1.7 & 2.0\\
  \hline
  $^7$Li    & 0.88 $\pm$ 0.03 & 11.6 $\pm$ 0.6 & 50.8 $\pm$ 2.6 & 0.36 $\pm$ 0.03 & 0.035 $\pm$ 0.003   & 20.9 $\pm$ 0.5 & 20.3 $\pm$ 2.6 & 3.1\\
  \hline
  $^8$Li    & 0.90 $\pm$ 0.09 & 11.9 $\pm$ 1.5 &  9.1 $\pm$ 1.4 & 0.45 $\pm$ 0.05 & 0.029 $\pm$ 0.005   & 18.0 $\pm$ 1.0 &  6.4 $\pm$ 1.5 & 2.1\\
  \hline
  $^9$Li    & 1.00 $\pm$ 0.22 & 10.4 $\pm$ 3.0 &  2.1 $\pm$ 0.5 & 0.39 $\pm$ 0.07 & 0.025 $\pm$ 0.003   & 18.2 $\pm$ 1.6 &  2.1 $\pm$ 0.6 & 1.2\\
%  \hline
  \hline
  $^7$Be    & 0.92 $\pm$ 0.27 & 11.2 $\pm$ 4.3 &  2.6 $\pm$ 0.8 & 0.48 $\pm$ 0.05 & 0.038 $\pm$ 0.005   & 24.0 $\pm$ 1.2 &  4.6 $\pm$ 0.9 & 1.4\\
  \hline
  $^9$Be    & 0.86 $\pm$ 0.12 &  9.6 $\pm$ 1.7 & 12.5 $\pm$ 1.9 & 0.53 $\pm$ 0.06 & 0.020 $\pm$ 0.005   & 16.6 $\pm$ 0.8 &  8.1 $\pm$ 2.3 & 1.4\\
  \hline
  $^{10}$Be & 0.90 $\pm$ 0.08 & 11.8 $\pm$ 1.2 & 10.0 $\pm$ 1.4 & 0.44 $\pm$ 0.04 & 0.026 $\pm$ 0.004   & 14.5 $\pm$ 0.9 &  6.8 $\pm$ 1.5 & 1.3\\
%  \hline
  \hline
  $^{10}$B  & 0.85 $\pm$ 0.20 & 10.5 $\pm$ 3.4 & 6.6  $\pm$ 1.3 & 0.73 $\pm$ 0.14 & 0.020 $\pm$ 0.010   & 18.2 $\pm$ 2.7 &  2.7 $\pm$ 1.7 & 1.8\\
  \hline
  $^{11}$B  & 0.93 $\pm$ 0.18 & 10.5 $\pm$ 2.1 & 12.8 $\pm$ 2.5 & 0.50 $\pm$ 0.05 & 0.022 $\pm$ 0.004   & 14.5 $\pm$ 0.7 & 12.8 $\pm$ 2.8 & 1.7\\
  \hline
  $^{12}$B  & 0.87            &  8.8           & 1.6            & 0.73            & 0.012               & 13.2           &  5.1           & 1.0\\
  \hline
\end{tabular}
%\end{sidewaystable}
\end{table*}

\begin{table*}
  \centering
  \caption{\label{table:elements}Parameters of one or two moving sources for elementally identified products }
\begin{tabular}{|l||l|l|l||l|l|l|l||l|}
  \hline
    & \multicolumn{3}{c||}{Slow source} & \multicolumn{4}{c||}{Fast source}                 &  \\
  \cline{2-8}
 Ejectile   & \emph{k}$_1$                & \emph{T}$_1$/MeV          & $\sigma_1$/mb & \emph{k}$_2$     &  $\beta_2$          & \emph{T}$_2$/MeV         & $\sigma_2$/mb & $\chi^2$ \\
  \hline
  \hline
  C         & 0.879             & 12.3           & 28.4             & 0.150             & 0.0367              & 15.8           & 11.8            & 3.34  \\
  \cline{2-9}
            &                   &                &                  & 0.759 $\pm$ 0.032 & 0.0076 $\pm$ 0.0007 & 13.4 $\pm$ 0.5 & 33.4 $\pm$ 0.9  & 3.26  \\
  \hline
  N         & 1.00              & 12.2           & 15.9             & 0.206             & 0.0382              & 14.5           & 9.6             & 1.47  \\
  \cline{2-9}
            &                   &                &                  & 0.822 $\pm$ 0.045 & 0.0118 $\pm$ 0.0007 & 13.4 $\pm$ 0.6 & 18.6 $\pm$ 0.7  & 1.41  \\
  \hline
  O         & 0.75              & 14.1           & 1.3              & 0.71             & 0.0129              & 12.2           & 11.8             & 0.78  \\
  \cline{2-9}
            &                   &                &                  & 0.784 $\pm$ 0.045 & 0.0118 $\pm$ 0.0007 & 13.5 $\pm$ 0.6 & 13.1 $\pm$ 0.6  & 0.75  \\
  \hline
  F         &                   &                &                  & 0.545 $\pm$ 0.079 & 0.0105 $\pm$ 0.0014 & 17.0 $\pm$ 3.2 & 5.12 $\pm$ 0.32 & 0.40 \\
  \hline
  Ne        &                   &                &                  & 0.670 $\pm$ 0.098 & 0.0090 $\pm$ 0.0018 & 15.3 $\pm$ 1.9 & 5.25 $\pm$ 0.44 & 0.54 \\
  \hline
  Na        &                   &                &                  & 0.801 $\pm$ 0.117 & 0.0104 $\pm$ 0.0020 & 12.0 $\pm$ 0.7 & 5.99 $\pm$ 0.63 & 0.61 \\
  \hline
  Mg        &                   &                &                  & 0.577 $\pm$ 0.090 & 0.0103 $\pm$ 0.0019 & 12.0 $\pm$ 1.2 & 4.59 $\pm$ 0.52 & 0.59  \\
  \hline
  Al        &                   &                &                  & 0.772 $\pm$ 0.180 & 0.0036 $\pm$ 0.0013 & 10.0 $\pm$ 0.7 & 5.24 $\pm$ 0.91 & 0.23 \\
 \hline
\end{tabular}
%\end{sidewaystable}
\end{table*}

\begin{figure}
\begin{center}
%\vspace*{-2cm}
\includegraphics[angle=0,width=0.5\textwidth]{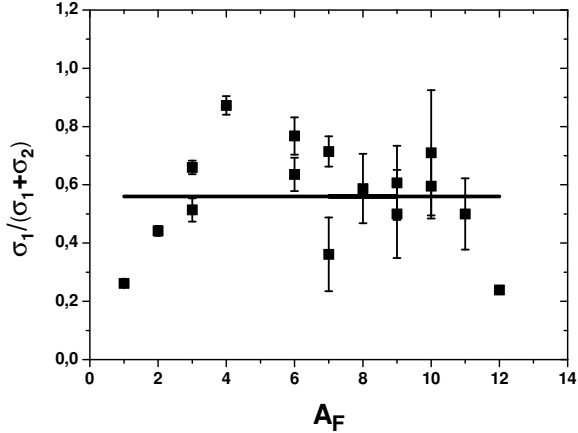}
\caption{\label{fig:s1tos12vsa} Ratio of total production cross
section corresponding to emission from the slow source to sum of the
total cross sections representing emission from both sources versus
mass number of the detected reaction product. The parameters $\sigma
_1$ and $\sigma _2$ were taken as total cross sections because
 they correspond to angle and energy integrated double differential
cross sections $d^2\sigma/d\Omega dE$.  The solid line shows average
value of the ratio (0.560 $\pm$ 0.044).}
\end{center}
\end{figure}

\begin{figure}
\begin{center}
\includegraphics[angle=0,width=0.5\textwidth]{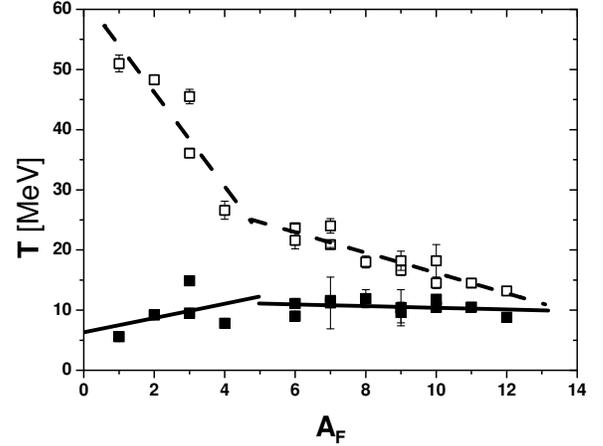}
\caption{\label{fig:t12vsa} Apparent temperature of the slow source
(full squares and solid lines) and that of the fast source (open
squares and dashed lines) versus mass number of the ejectiles. The
lines were fitted separately for light charged particles ($^1$H -
$^4$He) and intermediate mass fragments (A$_F \geq 6)$.}
\end{center}
\end{figure}

\begin{figure}
\begin{center}
%\vspace*{-2.5cm}
\includegraphics[angle=0,width=0.49\textwidth]{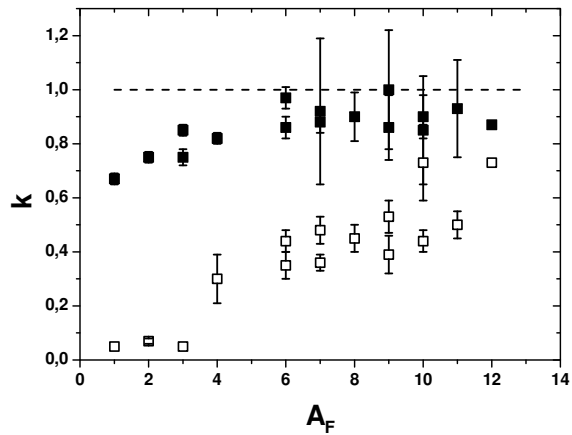}
\caption{\label{fig:k12vsa} The ejectile mass number dependence of
the factor scaling the Coulomb barrier of two touching spheres to
the actual height -- necessary for good description of the data.
Full squares represent the slow source and open squares show results
for the fast source. The line \emph{k}=1 is also depicted to
facilitate interpretation of the figure.}
\end{center}
\end{figure}

%  ----- beta of the fast source -----

\begin{figure}
\begin{center}
%\vspace*{-2.5cm}
%\includegraphics[angle=0,width=1.0\textwidth]{b2a.pdf}
\includegraphics[angle=0,width=0.51\textwidth]{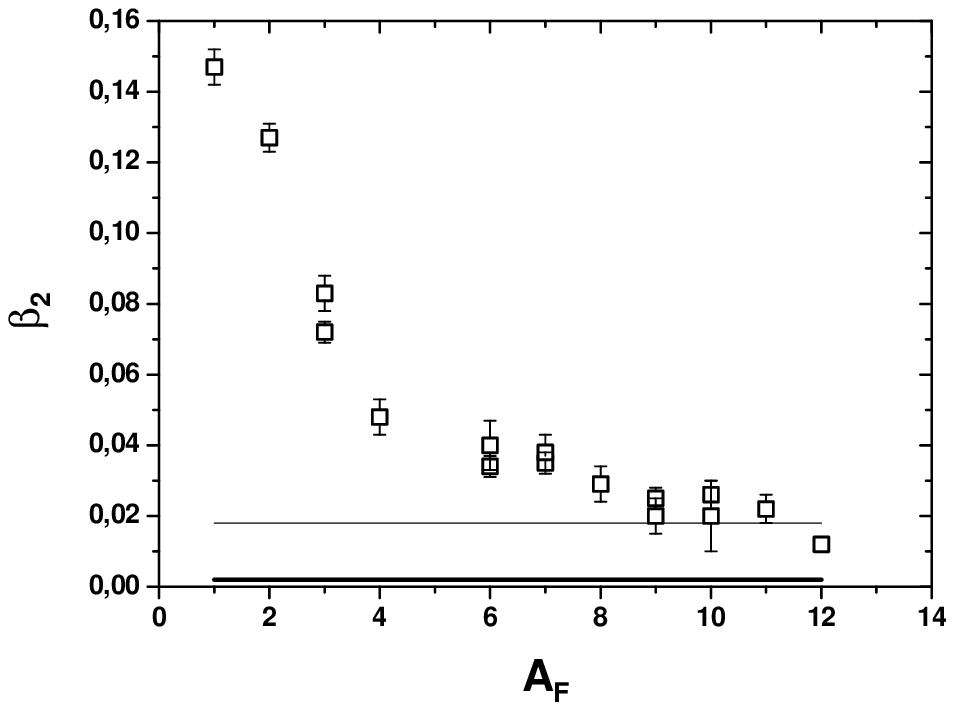}
\caption{\label{fig:b2vsa} Velocity of the second (fast) source as a
function of emitted fragment mass number - open squares. Thin solid
line $\beta_2$=0.018 represents velocity of the common center of
mass of the proton projectile and the target and thick solid line
$\beta_1$=0.002 shows velocity of the first (slow) source fixed at
velocity of heavy residuum of intranuclear cascade.}
\end{center}
\end{figure}

\bigskip
Thorough inspection of the parameter dependence leads to the
following conclusions:
\begin{enumerate}
  \item \textbf{\emph{The contribution $\sigma_1$ of the first (slow) emitting source is comparable
        to contribution $\sigma_2$ of the second source.}}

        It is illustrated by
        Fig. \ref{fig:s1tos12vsa} in which ratio of the total cross
        section for emission of ejectiles from the first
        source to the sum of the total cross sections for emission
        from both sources is shown as function of mass number of the
        ejectiles.  The average value of the ratio $\sigma_1/(\sigma_1+\sigma_2)$ is equal to 0.560
        $\pm$ 0.044.  It should be, however, emphasized that rather
        large deviations from the average value appear for
        individual ejectiles. For example, almost 90 \% of alpha particles
        is emitted from the slow source, whereas it is the case only  for $\sim$~25~\% of
        protons.

 \item \textbf{\emph{The parameters of the slow source have values which
       agree with the assumption that this source simulates evaporation from a heavy
       nucleus}} corresponding, \emph{e.g.}
       to the residuum of the target after the intranuclear cascade of
       nucleon-nucleon collisions, namely:
       \begin{itemize}
            \item The apparent temperature of the slow source is independent
                  of the mass of emitted intermediate mass
                  fragments, what can be seen in Fig.~\ref{fig:t12vsa}
                  where the slope parameter of the
                  solid line is equal to zero
                  within the limits of errors~:~-~0.15~$\pm$~0.17.
                  Its stability indicates that the
                  recoil effect of the source during emission of fragments is
                  negligible, thus, mass of the source
                  must be much larger than masses of observed IMF's (cf. Appendix).
                  Moreover, the horizontal line: \emph{T}~=~11.9~$\pm$~1.5~MeV,
                  fitted to temperature values extracted from spectra of IMF's
                  reproduces also quite well values of the apparent
                  temperature for light charged particles (H and He
                  ions) as can be expected for emission from heavy residuum
                  of the intranuclear cascade of nucleon-nucleon collisions.
            \item The $k$ parameter,  which determines the height of the effective
                  Coulomb barrier between the emitted fragment and the rest of
                  the emitting source (cf. Appendix) is very close to unity, what means that
                  the charge of the source does not differ significantly from
                  the charge of the target.
                  It is illustrated by Fig. \ref{fig:k12vsa} where
                  $k$ parameters for both sources are shown for
                  individual ejectiles. The full squares represent the slow source and
                  open squares correspond to the second, fast source.
                  The dashed line $k=1$ is shown in the figure to facilitate judgment on the
                  magnitude of the \emph{k} -- parameter.
       \end{itemize}
  \item \textbf{\emph{The second (fast) source is much lighter than the residual nucleus
        of the intranuclear cascade}} because:
        \begin{itemize}
          \item Its velocity is always larger
                than limiting velocity of the proton-target center of mass which
                would be obtained only when
                total beam momentum is transferred to the target ($\beta \approx
                0.018$).  Fig. \ref{fig:b2vsa} illustrates this fact
                showing $\beta_2$ values for individual ejectiles as
                well as the horizontal line $\beta=0.018$.
          \item The Coulomb barrier between the source and the
                ejectile is several times smaller than the
                Coulomb barrier of two touching charged spheres
                representing the target nucleus and the ejectile. This is well
                illustrated by open squares in Fig. \ref{fig:k12vsa}.
          \item The recoil effect (cf. Appendix) is clearly visible in the dependence
                of the apparent temperature of the source on the
                mass of the ejectile as it is shown in Fig.
                \ref{fig:t12vsa} - open squares.
        \end{itemize}

 \item     \textbf{\emph{The fast source describing  LCP's
      emission (hydrogen and helium ions up to $^4$He) is much lighter than the
      fast source responsible for emission of intermediate mass fragments.}}

      This may be inferred from different recoil effects visible as
      different slopes of two lines which describe the
      dependence of the apparent temperature on the mass of ejectiles
      (cf. Fig. \ref{fig:t12vsa}).
      The line corresponding to LCP's is more steep, giving
      the mass of the source equal to $A_S$=(8~$\pm$~2) nucleons, and very high
      temperature of the source $\tau$=(62~$\pm$~7) MeV (cf. Appendix for meaning of $\tau$).
      Velocity of this light source is very high;  $\beta~=~0.05~-~0.15$.
      Such a source can be, perhaps, identified with the
      \emph{fireball} created by the proton impinging on to the target together
      with nucleons present on its straight line way through the
      target nucleus \cite{WES76A},
% WES78A
% CUM80A 3 - 4 nucleons (effective target)

      The line describing temperature of IMF's corresponds to mass
      of the source $A_S$=(20 $\pm$ 3) nucleons and its temperature
      $\tau$= (33 $\pm$ 2) MeV.  Velocity of this source is much
      smaller  ($\beta= 0.02 - 0.04$) than velocity of source emitting LCP's.
%   \end{itemize}
\end{enumerate}

Very different properties of the fast source emitting light charged
particles (LCP's) and the fast source emitting intermediate mass
fragments (IMF's) leads to conclusion that the picture of two
sources is oversimplified. The presence of a fireball, which can
give contribution to emission of LCP's only and occurrence of the
light ($A_S \approx 20$), fast source emitting LCP's as well as
IMF's may be interpreted as indication  of a three body decay of the
target nucleus.  The third partner of such a decay would be heavy
and hardly distinguishable from heavy residuum of the intranuclear
cascade, therefore its occurrence could be described effectively by
the same slow source.

\section{Summary and conclusions}

The double differential cross sections $(d^2\sigma/d\Omega dE)$ were
for the first time measured with good statistics for isotopically
identified intermediate mass fragments produced by interaction of
2.5-GeV protons with the gold target.  The following individual
isotopes of the elements from hydrogen to boron were resolved:
$^{1,2,3}$H, $^{3,4,6}$He, $^{6,7,8,9}$Li, $^{7,9,10}$Be,
$^{10,11,12}$B, whereas for heavier ejectiles (from carbon to
aluminium) only elemental identification was done.
The energy spectra for all nuclear fragments, determined at several
scattering angles, appear to be of the Maxwellian shape with
exponential, high energy tail.  The low energy part of the
distribution is almost independent of angle, but the slope of high
energy tail of the spectrum increases monotonically with the angle.
The shape of the angle independent part of spectra can be reproduced
by the two-stage model of the reaction, i.e. intranuclear cascade of
the nucleon-nucleon and meson-nucleon collisions followed by
statistical emission from an equilibrated residual nucleus. However,
the absolute magnitude of the spectra predicted by two-stage model,
using Boltzmann-Uehling-Uhlenbeck program \cite{CAS06A} for the
intranuclear cascade and Generalized-Evaporation-Model (GEM)
\cite{FUR00A,FUR02A} for statistical emission of fragments, is in
agreement with the experimental data only for the light charged
particles (H and He ions). Furthermore, the theoretical cross
sections underestimate significantly the yield of heavier fragments
at high kinetic energies for all ejectiles. This indicates that
another mechanism plays
an important role besides the standard two-stage mechanism.\\

To get information on possible origin of this additional mechanism a
phenomenological analysis was performed assuming that the ejectiles
originate from two moving sources. The slow moving source was
identified with the heavy remnant nucleus of the first stage of the
two-step process mentioned above while no assumptions have been made
as concerns the second emitting source.  The properties of both
sources were treated as free parameters with exception of the
velocity of the slow source which was taken to be equal to the
velocity of the heavy residual nucleus from the intranuclear
cascade, namely
$\beta_1$=0.002.\\

Excellent agreement of the phenomenological parametrization with
experimental data was achieved with values of the parameters varying
smoothly from ejectile to ejectile.  Their behavior indicates that
the parameters of the slow source are compatible with the assumption
that it is a heavy nucleus which may be described as equilibrated
system of $\sim$ 12 MeV apparent temperature, whereas the second
source has completely different properties.  It corresponds either
to very small ($A_S \sim$ 8), very hot ($\tau \sim$ 62 MeV) and fast
($\beta= 0.05 - 0.15$) fireball or to heavier ($A_S \sim 20 $),
colder ($ \tau \sim$ 33 MeV), and slower ($\beta= 0.02 - 0.04$)
cluster.\\

The described above properties of two emitting sources observed in
the interaction of 2.5 GeV protons with the gold target lead to a
conclusion that two different mechanisms are observed giving almost
the same contribution to the cross sections. First of them is
compatible with the standard, two-stage model whereas another one
seems to be similar to the picture of cold break-up proposed by J.
Aichelin, J.H\"ufner and R. Ibarra
\cite{AIC84A}.\\

 In the model of cold break-up the energetic proton bombarding the target
drills a cylindrical hole through the nucleus causing that the
deformed remnant of the collision breaks up into two pieces.  Thus,
three correlated groups of nucleons appear after first, short stage
of the reaction : (i) the fast, small cluster consisted of the
nucleons which were placed within the cylinder with the axis along
to the projectile path,(ii)  two clusters - products of the break
up. All three clusters act as sources emitting light charged
particles, whereas two heavier clusters are also responsible for
emission of intermediate mass fragments.  The two latter clusters
are produced in result of a dynamical process in which the "wounded"
nucleus cannot come to its ground state by emission of nucleons or
small clusters, however, the correlation of the fast group of
nucleons knocked out by the projectile is of another, kinematic
origin. The high energy proton impinging on to the nucleus sees it -
due to Lorentz contraction - as a narrow disc. Therefore all the
nucleons which lie on the path of the projectile interact
simultaneously, as one entity, with the projectile. It was shown
that this collectivity affects the multi particle production in
proton-nucleus collisions \cite{BER79A} as well as manifests itself
in enhanced dependence of momentum transfer on projectile energy in
deep-spallation reactions \cite{CUM80A}. Thus, it is not surprising
that such correlated group of nucleons can appear as a hot, fast
moving source emitting light charged particles  observed in the
present experiment.  Of course, it cannot give contribution to
emission of intermediate mass fragments because the source is
consisted of a few nucleons only.\\

It might be something confusing why in the present parametrization
only two sources were necessary for good description of the data if
the postulated cold break-up mechanism of the reaction calls for
presence of three sources.  This apparent inconsistency is easy to
be removed: The slow, heavy source represents the heavy residual
nucleus produced by the standard two-step model of the reaction
and/or the heavy fragment from the break up . The light, fast source
is responsible for simulation of the hot fireball (for light charged
particles) or the lighter fragment from the break
up (for intermediate mass fragments).\\

Consistency of the present phenomenological description with the
cold break up picture of the reaction cannot be assumed as a proof
of the underlying reaction  mechanism. Additional experimental facts
should be searched for establishing the confidence in such an
interpretation. For example, it can be expected that similar
phenomena have to appear for other heavy target nuclei if they are
observed for the gold target. On the contrary, it is not obvious
whether proton induced reactions on light targets or at
significantly different proton energies should show similar
behavior.

\appendix*
\section{\label{appendix} Phenomenological parametrization}

In this Appendix assumptions and details of the formulation of two
moving source model are discussed.  The content of the appendix is
very close to information contained in the original paper of
Westfall \emph{et al.} \cite{WES78A}, however, the additional
modification and properties introduced in the model need
to be discussed for proper understanding of the performed analysis.\\

The model assumes that the nucleons and composite particles are
emitted from two moving sources with the following properties:

\begin{description}
\item[(i)] Each source moves along the proton beam direction,

\item[(ii)] Angular distribution of emitted particles is isotropic
in the source rest frame,

\item[(iii)] Distribution of the kinetic energy $E^*$ available in the two-body
break up of the source has a Maxwellian shape characterized by the
temperature parameter $\tau$:
 \[
\frac{{d^2 \sigma }}{{dE^* d\Omega ^* }} = \frac{\sigma }{{2\left(
{\pi \tau } \right)^{3/2} }}\sqrt {E^* } \exp \left[ { - \frac{{E^*
}}{\tau }} \right].
\]
\end{description}

\noindent The distribution is normalized in such a way that
integration over angles and energies gives the total cross section
equal to the
parameter $\sigma$.\\

Since the mass of the source $A_S$ is finite, the energy and
momentum conservation laws cause that the energy $E^{'}$ of the
observed particle of mass $A_F$ differs from the full kinetic energy
$E^*$ available in the source frame:

\[
 E^{*}  =  \nu E^{'}   \;\;\; {\rm where} \;\;\;
 \nu   \equiv  \frac{A_S }{A_S  - A_F},
\]

\noindent thus the energy distribution of the emitted fragment in
the rest frame of the source is given by:
\[
\frac{{d^2 \sigma }}{{dE^{'} d\Omega ^{'} }} = \frac{{\nu \sigma
}}{{2\left( {\pi \tau } \right)^{3/2} }}\sqrt {\nu  E^{'} } \exp
\left[ { - \frac{{\nu E^{'} }}{\tau }} \right].
\]

\noindent This formula is usually applied without explicit writing
the recoil correction, i.e. by introducing so called \emph{apparent
temperature} $T \equiv \tau / \nu$:

\[
\frac{{d^2 \sigma }}{{dE^{'} d\Omega ^{'} }} = \frac{{ \sigma
}}{{2\left( {\pi T} \right)^{3/2} }}\sqrt {  E^{'} } \exp \left[ { -
\frac{{E^{'} }}{T }} \right].
\]

\noindent Such a form of this formula is used also in the present
paper.

\vspace*{0.5cm} It is worth to note, that the recoil of the source
gives a possibility to extract the information on the source
temperature parameter $\tau$ as well as on the mass of the source
$A_S$ from linear dependence of the apparent temperature $T$ on the
fragment mass $A_F$:

\[
T \equiv \frac{\tau }{\nu }  = \tau  - \left( {\frac{\tau }{{A_S }}}
\right) \cdot A_F \;.
\]

The charged particles emitted from the source must overcome the
Coulomb barrier, what significantly changes the shape of the low
energy part of their spectrum.  The presence of the barrier may be
taken into account by shifting the argument in the Maxwell formula
by the height of the barrier, as it was originally proposed in Ref.
\cite{WES78A} or by multiplying the Maxwell distribution by the
transmission factor. The first method is equivalent to the
application of a sharp cut-off what is a too crude approximation,
thus the result must be averaged over some distribution of heights
of the barriers \cite{WES78A}. The second method explicitly
introduces a smooth variation of the transition probability with
energy, however, this method also must introduce some assumptions
concerning height and curvature of the barrier.  In the present
work, the probability $P$ to overcome the Coulomb barrier was
parameterized in the following form:

\[
P = \frac{1}{1 + \exp \left[ - \left(\frac{E - k \cdot B}{d}\right)
\right]} \;,
\]

\noindent where $B$ is the Coulomb barrier of two touching spheres
corresponding to the emitted fragment of mass number $A_F$ and
charge number $Z_F$ and to the remaining part of the source with the
mass number of $(A_S -A_F)$ and charge number $(Z_S-Z_F)$:

\[
B=\frac{Z_F(Z_S-Z_F)e^2}{1.44\left({A_F}^{1/3}+(A_S -
A_F)^{1/3}\right)} \;.
\]

The quantities \emph{k} and \emph{d} are the parameters.  The first
parameter (\emph{k}) gives magnitude of the Coulomb barrier in units
of $B$.  To avoid ambiguity of $B$ determination arising from the
fact that at least two different sources are present in the current
analysis, we evaluated $B$ value assuming that $Z_S=79$ and
$A_S=197$, \emph{i.e.} there are atomic and mass numbers of the
target. Such value of $B$ is a good approximation of the Coulomb
barrier for heavy residua of the intranuclear cascade, thus one
should expect that then the $k$ parameter is close to unity.
However, with such definition of $B$, the $k$ parameter should be
significantly smaller than unity for light sources. The parameter
$k$ was searched by looking for the best fit of model spectra to the
experimental data.
The second parameter (\emph{d}) was arbitrarily fixed in the present
analysis by keeping constant the ratio of the height of the barrier
\emph{kB} to its diffuseness parameter \emph{d}: $kB/d=5.5$.

The explicit introduction of the barrier penetration factor $P$
gives finally the following formula for the double differential
cross section $d^2\sigma /dE^{'} d\Omega^{'} $:

\begin{eqnarray*}
 \frac{d^2\sigma }{dE^{'} d\Omega^{'} } &=& \frac{\sigma }{4\pi T^{3/2} I(kB,d,T)} \cdot \frac{\sqrt{E^{'}} \exp\left( {- \frac{E^{'}}{T}} \right)}{1 + \exp \left( \frac{kB - E^{'}}{d} \right)} \\
 I(B,d,T) &=& \int\limits_0^\infty  {\frac{{dx \cdot \sqrt x  \cdot \exp \left( { - x} \right)}}{{1 + \exp \left( \frac{kB- T \cdot x}{d} \right)}}}
\end{eqnarray*}

\noindent The integral $I(B,d,T)$ used for normalization  of the
distribution (preserving previous interpretation of $\sigma$
parameter) has been evaluated numerically by the Gauss-Laguerre
method.

It is necessary to transform the model double differential cross
sections calculated in the rest frame of the emitting source to the
laboratory system when comparing the model predictions to
experimental data. It can be shown that the transformation may be
performed by following formula:

\[
\frac{{d^2\sigma }}{{dEd\Omega }} = \frac{p}{{p^{'} }} \cdot
\frac{{d^2\sigma }}{{dE^{'} d\Omega ^{'} }} \approx \sqrt
{\frac{E}{{E^{'} }}}  \cdot \frac{{d^2\sigma }}{{dE^{'} d\Omega ^{'}
}}\;,
\]

\noindent where the first equality is exact and the second is valid
in nonrelativistic limit, normally realized in the motion of
observed ejectiles.  The nonrelativistic relationship between
kinetic energy $E$ of the particle emitted at the angle
$\theta_{LAB}$ in the laboratory system and the energy $E^{'}$ of
emitted particle  in the rest frame of the source is as follows:
\[
E^{'}  = E + \frac{{m\beta ^2 }}{2} - \sqrt {2mE}  \cdot \beta \cdot
\cos \theta _{LAB}\;,
\]
\noindent where $m$ ($\equiv A_F$) is the mass of the emitted
particle and $\beta$ - the
velocity of the source in the laboratory system.\\

%\section{}

\begin{acknowledgments}
The quality of the beam necessary for the success of this work is due mainly to the efforts of the COSY operator crew. The authors acknowledge gratefully the support of the European Community-Research Infrastructure Activity under FP6 "Structuring the European Research Area" programme (CARE-BENE, contract number RII3-CT-2003-506395 and HadronPhysics, contract number RII3-CT-2004-506078). The authors appreciate the financial support of the European Commission through the FP6 IP-EUROTRANS FI6W-CT-2004-516520.
\end{acknowledgments}

\bibliography{spallit}

\end{document}